\documentclass[journal]{IEEEtran}
\usepackage{amssymb}
\usepackage[T1]{fontenc}
\usepackage{placeins}
\usepackage{booktabs}
\usepackage{graphicx}
\usepackage{hyperref}
\usepackage{amssymb,amsmath,bm}
\usepackage{textcomp}
\usepackage{graphicx}
\usepackage{wrapfig}
\usepackage{lipsum}
\usepackage{dblfloatfix}
\usepackage{fixltx2e}
\usepackage{nameref}
\usepackage{amsmath}
\usepackage{hyperref}
\usepackage{graphicx}
\usepackage{varioref}
\usepackage{booktabs}  
\usepackage{siunitx}
\usepackage{numprint}
\usepackage{textcomp}

\usepackage{lipsum}
\ifCLASSOPTIONcompsoc
  \usepackage[nocompress]{cite}
\else
  \usepackage{cite}
\fi
\ifCLASSINFOpdf
\else
\fi
\begin{document}
%
\title{Fog-Assisted wIoT: A Smart Fog Gateway for End-to-End Analytics\\ in Wearable Internet of Things}
\author{\IEEEauthorblockN{Nicholas Constant\textsuperscript{*}, Debanjan Borthakur\textsuperscript{*}, Mohammadreza Abtahi, Harishchandra Dubey, Kunal Mankodiya\thanks{\textbf{This material is presented to ensure timely dissemination of scholarly and technical work. Copyright and all rights therein are retained by the authors or by the respective copyright holders. The original citation of this paper is: N. Constant , D. Borthakur, M. Abtahi, H. Dubey, K. Mankodiya, "Fog-Assisted wIoT: A Smart Fog Gateway for End-to-End Analytics in Wearable Internet of Things", \textit{The 23rd IEEE Symposium on High Performance Computer Architecture HPCA 2017}, (Feb. 4, 2017 – Feb. 8, 2017), Austin, Texas, USA.}}}
\IEEEauthorblockA{Department of Electrical, Computer and Biomedical Engineering, University of Rhode Island, RI, USA\\
Email: kabuki4774@my.uri.edu, kunalm@uri.edu\\
\textsuperscript{*}Authors contributed equally}
}
\maketitle
\begin{abstract}
Today, wearable internet-of-things (wIoT) devices continuously flood the cloud data centers at an enormous rate. This increases a demand to deploy an edge infrastructure for computing, intelligence, and storage close to the users. The emerging paradigm of fog computing could play an important role to make wIoT more efficient and affordable. Fog computing is known as the cloud on the ground. This paper presents an end-to-end architecture that performs data conditioning and intelligent filtering for generating smart analytics from wearable data. In wIoT, wearable sensor devices serve on one end while the cloud backend offers services on the other end. We developed a prototype of smart fog gateway (a middle layer) using Intel Edison and  Raspberry Pi. We discussed the role of the smart fog gateway in orchestrating the process of data conditioning, intelligent filtering, smart analytics, and selective transfer to the cloud for long-term storage and temporal variability monitoring. We benchmarked the performance of developed prototypes on real-world data from smart e-textile gloves. Results demonstrated the usability and potential of proposed architecture for converting the real-world data into useful analytics while making use of knowledge-based models. In this way, the smart fog gateway enhances the end-to-end interaction between wearables (sensor devices) and the cloud. 
\end{abstract}
%
\IEEEpeerreviewmaketitle
\section{Introduction}
%
\begin{figure*}[!tb]
\centering
\includegraphics[width=450pt]{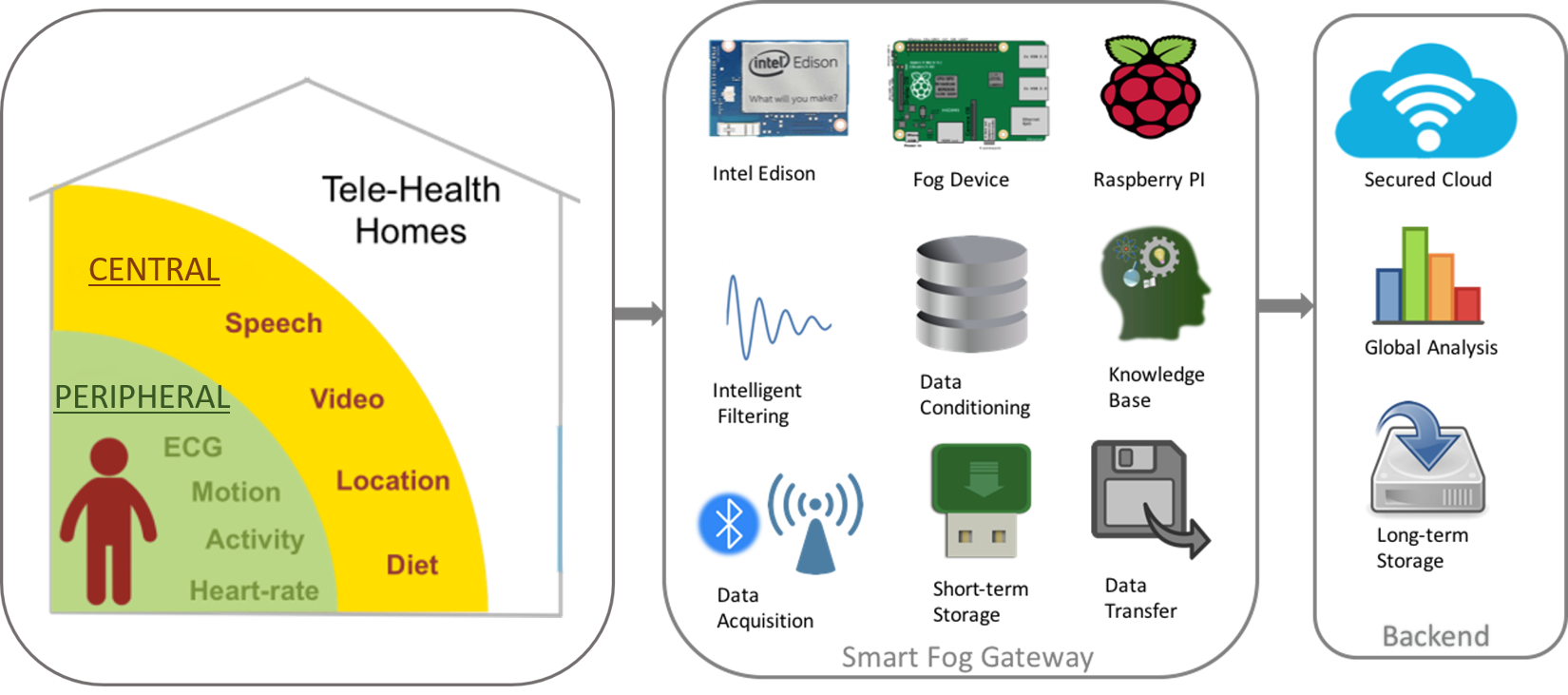}
\caption{The conceptual overview of the fog-assisted wIoT architecture for end (sensors)-to-end(cloud) analytics. The data generated in wearables is conditioned within fog device. In addition, temporary storage and intelligent filtering, analysis and prediction is performed on data. The final analysis were sent to cloud backend for long-term storage and global analysis.}
\label{fig_concept}
\end{figure*}
Wearable sensors continuously collect and stream data to cloud devices for data processing and long-term storage. These wearable sensors are low-power, low-bandwidth devices, and close proximity devices. In an effort to expand battery-life and response time of the sensors, the fog concept, often referred to as edge computing, has grown more common-place among system architectures. The fog concept coined by CISCO aims to minimize latency, conserve bandwidth, improve security, maintain reliability, and move data to the best place for processing~\cite{r1}.  
This concept allows for more sensors, or edge devices, to interface with the cloud on a much larger scale than previously seen. Since the cloud is not setup for this volume and variety of data changes in the systems before the cloud must be adapted to better utilize the cloud services available. That is why in this paper we will discuss the development of a smart fog to help mitigate the amount of data sent to the cloud by playing the role of orchestrator in the process of data acquisition, conditioning, analysis, and short-term storage. Our discussion will start with an overview of the fog environment by defining the roles of an IoT device, fog node, and cloud. This will transition into a description of the communication protocols used between the groups, along with the fog setup used during the benchmarking. Finally, we compare the results of our benchmarking experiments to those of realistic payloads to explore limitations in scale.
%
\vspace{-2mm}
\section{Related Works}
\vspace{-2mm}
\subsection{Wearable Internet-of-Things - wIoT}
\vspace{-1mm}
IoT or internet of things is the internet working of physical devices or other items embedded with electronics, software, sensors and network connectivity. IoT enables the connected objects to collect and exchange data \cite{itu}. We define a new segment of IoT that is named as wearable IoT (wIoT) \cite{hiremath2014wearable}. Wearable devices are rapidly emerging and this gives rise to the wIoT due to their capability of sensing, computing and communication \cite{hiremath2014wearable}. We also discussed the building blocks of wIoT which includes wearable sensors, internet-connected gateways and cloud and big data support that can be instrumental in its future success in healthcare domain applications \cite{hiremath2014wearable}.
IoT is transforming lives. Wearable devices that offer biometric measurements such as heart rate, perspiration levels, oxygen levels in the bloodstream are available even in the smartwatch itself. In \cite{new} authors assert that technology advancements may even allow alcohol levels or other similar measurements to be made via a wearable device. Complex sensors are increasing the efficacy of wIoT, like tracking body temperature might provide an indication of flu. Wearable devices connectivity with household appliances can make its work more versatile. WIoT light sensors might be used for controlling the lighting in home etc. The privacy threat that might come into picture while dealing with biometrics associated with wIoT can be an interesting issue to tackle in near future. Nobody wants to share their blood pressure levels to an unknown person, unless he or she is their doctor.
For smooth working of high-end IoT applications, significant processing and interface capabilities are always needed.  

\subsection{Management of Big Data in Wearable Internet of Things}
Widespread use of wearable and internet of things have lead to several interesting applications and also unprecedented challenges. For enhancing the data management and analytics in wIoT, edge computing have emerged. Fog computing is emerging as a tool for such situations. The data generated by wearables are of temporal and spatial nature. Employing edge devices for analysis and visualization of data leads to efficient solution leading to improvement in overall power efficiency. The big data being generated from various applications could be explained by four Vs namely volume, velocity, variety 
and veracity~\cite{sun2016internet}. The harmony between wIoT and big data could lead to generating valuable analytics from big data. Fog computing holds a great promise to reduce the burden of wearable big data at the edge of the network. Authors in \cite{dubey2016bigear} propose a novel BigEAR big data framework, where it identifies mood of the person from various activities such as laughing, singing, crying, arguing and sighing. Our proposed Fog gateway can be made to incorporate such versatile clinical speech processing framework. Such framework is well accounted in literature such as in\cite{monteiro2016fit}, where authors present a Fog computing interface \cite{monteiro2016fit} for processing clinical speech data. 

\subsection{Fog Computing: Architecture, Feasibility and Opportunities}

Fog computing as defined in \cite{itu1} is a model to complement
the cloud for decentralizing the concentration of computing
resources (for example, servers, storage, applications and
services) in data centers towards users for improving the
quality of service and their experience. Fog computing \cite{itu1} is the process of decentralization of computations which is away from the cloud and towards the edge of the network closer to the user. In this way, fog computing increases the quality of service and reduces the latency and frequency of communication between a user and an edge node. Moreover, fog computing can improve the efficiency and performance of application \cite{itu}. In\cite{meng2017two} authors emphasize on the increasing need for implementing machine learning algorithms including deep learning on resource constrained mobile embedded devices with limited memory and computing power. Authors used a $2$ Bit network for model compression, this achieves a good trade-off between model size and performance. The scopes and opportunities of Fog computing is enumerable. Fog computing supports a growing variety of applications such as those in the Internet of Things(IoT), Fifth generation (5G) wireless systems, and embedded Artificial Intelligence (AI) \cite{chiang2016fog}.
\vspace{-2mm}
\section{The Architecture of Smart Fog Gateway}
\vspace{-2mm}
The fog environment consists of IoT devices, fog nodes, and the cloud. The fog is aimed at alleviating strain on the cloud by sharing summarized activities of the IoT devices, as well as enacting the learned rules created on the cloud. In order to develop the context in which the fog node is applicable, the key roles for the groups identified in Figure 1 will be explained.

\subsection{Smart Fog Computing: Semantics, Cognition and Perception}
We used knowledge-based models for computation and analytics from big data collected by wearables connected to internet.
Smart IoT-Based Intelligent Perception is emerging as a new service-oriented model. Authors in \cite{tao2014iot} propose a five-layered
structure  (i.e., resource layer, perception layer, network layer, service layer, and application layer) resource intelligent perception
and access system based on IoT. It not only provides promising attention to healthcare but also in powerful industrial systems utilizing the ever growing wireless, sensor devices and embedded systems. Social IoT came which encompasses social networks with IoT. Intelligent perception and making use of 
different sensing devices and adapters Fog actually mimics human perception.
Sensing devices might include barcode,
RFID readers, sensors, GPS, etc. The
adapter includes software interface adapter, sensor adapters,
model adapters, etc. Smart Fog devices might be well configured to be adaptive to real-time data streams. They can be modeled with algorithms that will make their processing energy efficient, encrypted and with low latency, and hence the name 'smart' aptly befits the Fog gateway proposed in this paper.

%
\subsection{IoT Device}
The IoT devices interacting with the fog node will minimally consist of sensors or actuators capable of transmitting collected data via Bluetooth. These devices may be limited in power, processing and storage. The topology among the devices can vary, but in this paper we focus on
a mesh topology. The main role of the IoT device is to sample real world data, when needed transmit that data to the fog node for decisions and storage. These devices may sample at high rates for applications such as biological sensing, down to low rates for applications such as monitoring contents within a fridge.
\subsection{Fog Node}
The role for a fog node is to orchestrate communication with the IoT devices and deliver real-time responses based off of rules determined by the supervised learning. The fog node will condition, analyze, and package this data for the cloud to learn and store. The fog node can use a
variety of software for the cleaning and learning, but this paper will focus on the use of Octave. The nodes and IoT devices are configured in a self-healing mesh topology using a routing technique. Communication to the cloud is accomplished via web-sockets. This paper focuses on the use of Socket.io for Node.js.
\subsection{Cloud}
The cloud acts as the overseer for the overall system by employing a combination of deep learning, long-term storage, and more. However, this paper focuses on the use of TensorFlow for deep learning. The cloud can disperse the latest updates down to the IoT device via the fog node.
The cloud is not required to deliver real-time responses down to the fog. Instead, the intention is to migrate that responsibility to the fog node.
\subsection{Communication Protocols Used}
\begin{enumerate}
\item Bluetooth $4$: Between IoT devices and Fog Nodes
\item Trickle Algorithm: Workaround to create mesh topology using Bluetooth $4$
\item Socket.io: Between Fog Nodes and Cloud
\end{enumerate}

A key component to in a fog environment is the ability for continuous communication in
directions down to the IoT devices and up to the cloud. There are currently a variety of wireless
protocols based off of the $IEEE 802.15.4$ such as Thread, ZigBee, Bluetooth, and more. Each of
these protocols allow for mesh topologies. Although we will focus primarily on the use of
Bluetooth among IoT devices and fog nodes. The communication within IoT devices needs to be consistent over a long range while maintaining low power. The bandwidth required for the various devices depends on the application, although our wearables are satisfied with the far less than 25Mbps as provided by Bluetooth $4$. 
\vspace{-2mm}
\section{Results \& Discussions}
\vspace{-2mm}
\subsection{System Description}
\vspace{-1mm}
The Intel Edison platform used in this analysis has a dual-core, dual-threaded Intel Atom CPU at 500MHz with a  32-bit Intel Quark microcontroller working at 100MHz, Bluetooth $4.0$ and dual-band $IEEE 802.11a/b/g/n$ via an on-board chip antenna. Yocto is the Linux environment for this plateform, a prebuilt distribution of Debian/Jessie for 32-bit systems was deployed as Yocto is not an embedded distribution of Linux itself. It only provides an environment to develop a custom Linux distribution. This was done so that same environment on both the Intel Edison and the Raspberry Pi can be achieved.

The Raspberry Pi Model B platform used in this analysis has a system consisting of a 900MHz 32-bit quad-core ARM Cortex-A7 CPU, with 1GB RAM. A WIFI dongle based on the Realtek $RTL8188CUS$ chipset was installed as Raspberry Pi does not have WIFI connectivity. Raspbian is the custom Linux distribution for this platform. Raspbian was replaced with the Debian/Jessie distribution used on the Intel Edison for computational flexibility.
\subsection{Smart Glove Data}
\begin{figure}[h!]
\includegraphics[width=250pt]{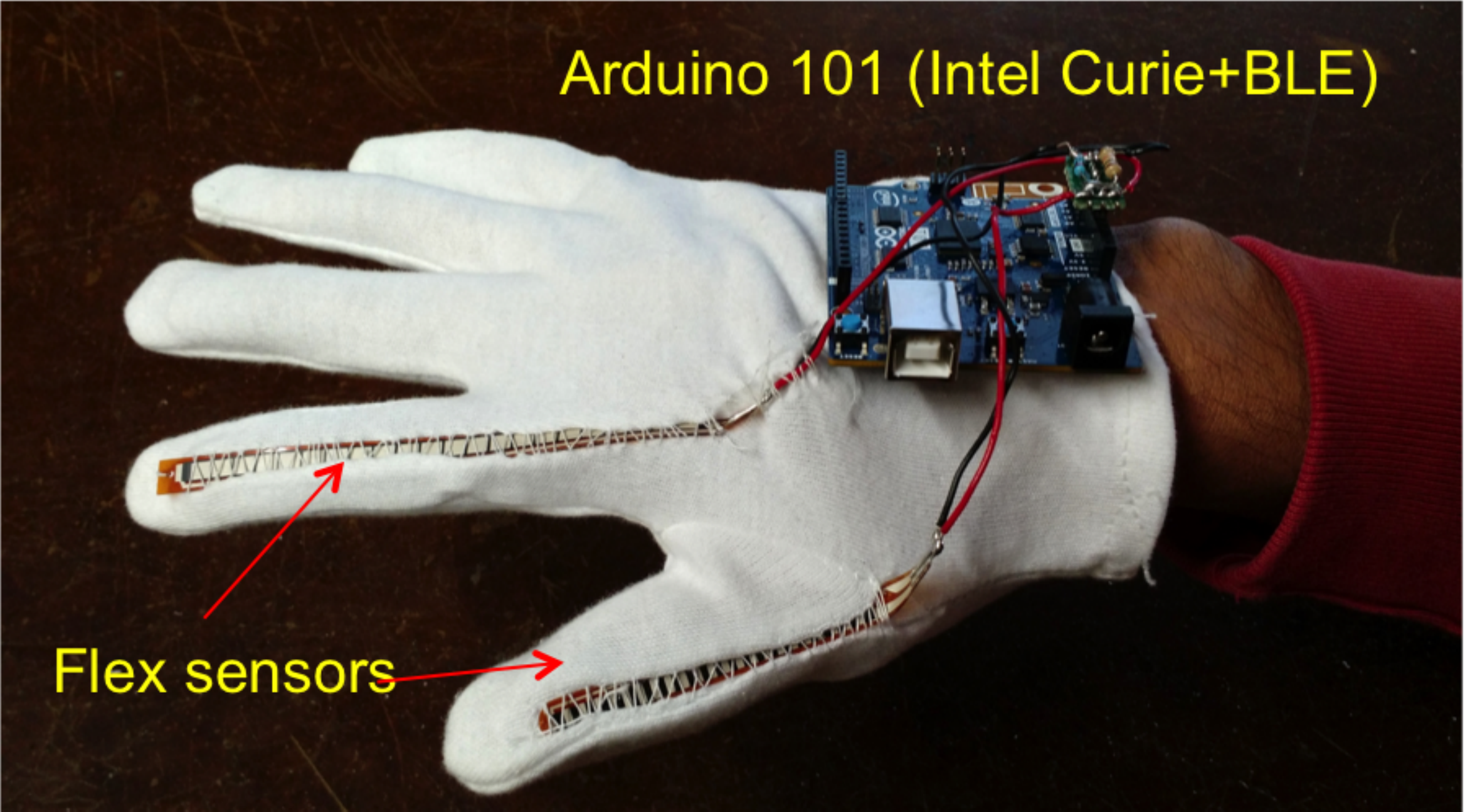}
\caption{Illustration of Smart Glove showing the microcontroller Intel Curie and the location of flex sensors.}
\label{glove}
\end{figure}

\begin{figure}[h!]
\includegraphics[width=250pt]{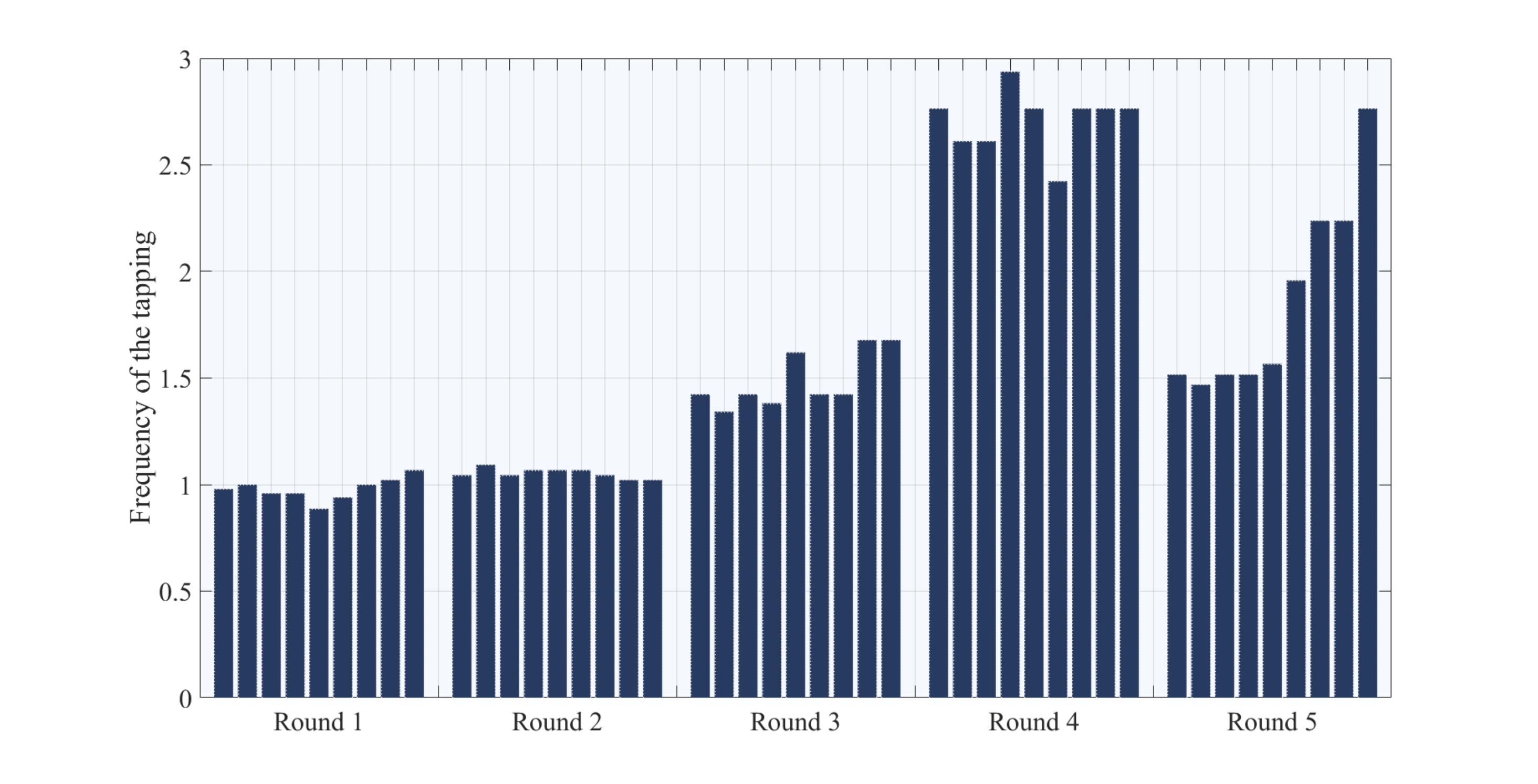}
\caption{Frequency of finger tapping resulted from the Smart Glove. Rounds 1 and 2 are tapping slowly, round 3 getting faster and round 4 is as fast as possible, round 5 starts with tapping slowly and gradually getting faster.}
\label{freq}
\end{figure}

Smart Glove\cite{gloves2017} is a wearable, completely wireless device transmitting the data recorded by the microcontroller Arduino $101$ to a smart phone or computer via Bluetooth. The sensors that are integrated in the Smart Glove are Spectra Symbol flex sensors that have a thickness of $6.35mm$, with $84.86\%$ of the part length designated as active length. The flex sensors are analog resistors which act as variable analog voltage dividers. The voltage across the flex sensors will change if they are bent, therefore, we can convert it to the resistance of the flex sensor that changes, and the aim is to measure the angular displacements where the flex sensor is located. 
In our prototype design of the Smart Glove, we have used two flex sensors located on pointing finger and thumb (see Figure \ref{glove} in order to quantify how much these fingers are bent during the experiment of finger tapping which is a common practice in the treatment procedure of Parkinson's disease. 
In this experiment, the Smart Glove data can reveal how much the participant has bent their fingers to do the finger tapping, and also show how fast they can do it. Figure \ref{freq} shows the frequency of finger tapping in an experiment that the participant was asked to start finger tapping slowly, and gradually doing it faster at different rounds, and in the last round, were asked to start slowly and getting it faster.
Therefore, the Smart Glove can be used as a wIoT device in the movement therapy of any kind of disorders that needs monitoring of the hand motion, such as Parkinson's disease. The patients can wear the Smart Glove at home, doing the assigned exercises and the data can be sent to the physician, without having to make a trip to visit the doctor.
\vspace{-2mm}
%
%
%
\vspace{-2mm}
\subsection{Benchmarking and Program Setup}
\vspace{-1mm}
The Fog devices were logged into with the SSH protocol. Some Benchmarking scripts were then run. The scripts starts Octave and load it with the data. The script searches for  the process ID (PID). Once determined the PID, it would extract all the information top provided about the systems performance and also the load on the system by this instance of Octave. The information was logged into a $.csv$ file and saved for analysis. Profiler function in the background starts once the instance of Octave was ready to run the algorithm. At the end, the Profilers set of data was stored into a $.mat$ file for later analysis.
\vspace{-2mm}
\subsection{Analysis}
\begin{figure}[h!]
\includegraphics[width=250pt]{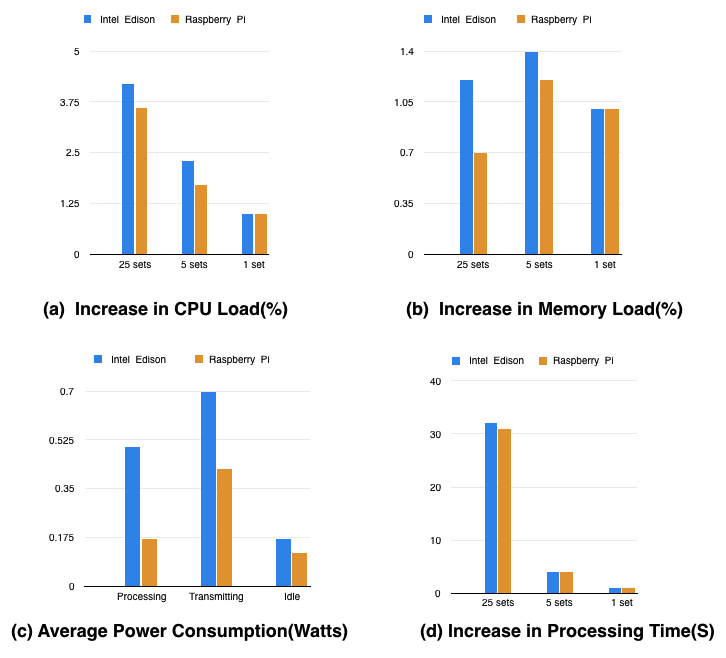}

\caption{Performance comparison between Intel Edison and Raspberry Pi in proposed architecture. We used real-world smart-glove data for this benchmarking experiment.The average load put onto the board when processing is shown here. Also Memory Load and CPU load are shown as percentages based on  data-set.}

\label{fig_perform}
\end{figure}

\vspace{-1mm}
The total process breakdown shown 
in Figure \ref{fig_perform} includes the load added on by starting an instance of Octave. The run  time for the Intel
Edison was much greater than that of the Raspberry Pi and an increase in data sets (size N) produces a processing time of the order of Nlog(N). It is noticed that  Raspberry Pi could complete the process almost two times faster and it could scale to 125+ datasets where as the Edison might crash. The network created by us places the Fog server in between the Smart Glove (or other wearables) and the cloud. It was assumed that the mean arrival rate of data would be once per minute assuming  the Fog device would be placed in locations where only a small number of devices in the area exist, such as a hospital. The service rate for the Fog servers could be summarized  from Figure 3 under total run-time. Little's Law can be used to  determine the average wait time for each device, that is\\
$Lead Time = WIP (units) / ACR (units per time period)$,\\ where WIP is Work-in-Process and ACR is Average Completion Rate  for any process \cite{isixsigma}.

While running  one set at a time, the average wait time for the Intel Edison and Raspberry Pi was around $64.65$ seconds and $12.39$ seconds, respectively. Both of the  devices used as fog servers were used to collect data from the smart glove, smartwatch, wristband and/or other wearable devices, then next step was data processing and finally, the processed data was sent to a server in the cloud. Figure \ref{fig_perform} shows the average amount of time spent on each step. The Raspberry Pi provided a service rate that is one-third of that of Intel Edison. Also, it was found Raspberry Pi consumes $198 mW/s$ and the Edison consumes $529 mW/s$ when active.

%

%

%

\subsection{Limitations}
Cloud and Fog always play a mutually beneficial inter-dependent relationship. The limitations might encompass several areas, one of them is the Cloud-Fog Interface. Fog can not provide the massive storage and complex heavy computations and the problem of wide area connectivity is another issue. Another limitation may be the security. Authors in \cite{chiang2016fog} mention that, though fog might enhance security but it might pose some new security challenges while dealing with privacy
sensitive data. Other privacy issues that might include data privacy, location privacy, usage privacy i.e. the usage pattern with which a fog client utilize the fog services. In the paper \cite{yi2015security} authors give an example of it, in smart grid, the reading of the smart meter will disclose lots of information of a household, such as at what time there is no person at home, and at what time the TV is turned on, which absolutely breaches user's privacy. Inadequate storage and inability to do high performance computing is another area of concern for fog devices.
The research presented in this paper utilized a clunky mesh network for the IoT devices. In future works, Bluetooth $5$ will be used which will provide a proper mesh network for these devices.
%


\section{Conclusion}

Massive amounts of data is being generated with widespread use of wearables, sensor devices in a variety of applications such as healthcare, smart-city, smart-grid, assisted homes~\emph{etc.}. We proposed, developed and validated a smart fog gateway that could perform end-to-end analytics for internet-connected wearables. Comparison between Intel Edison and Raspberry Pi provide evidence on choosing embedded boards that could match the demands of the application. The benchmarking experiments were done on real-world data. 

\nocite{*}
\bibliographystyle{IEEEtran}
\bibliography{ref}
\end{document}